\documentstyle[12pt]{article}
\begin{document}
\baselineskip 19pt plus 2pt
\renewcommand{\thesection}{\arabic{section}}
\renewcommand{\theequation}{\thesection.\arabic{equation}}

\hspace{10.5cm}
BRX-TH-350
           \vspace{0.4cm}
           \begin{center}
           \begin{bf}
           \begin{Large}
Quantum Dust Black Holes \\
           \end{Large}
\vspace{0.4cm}
A Statistical Mechanics Derivation \\
of Black Holes Thermodynamics \\
            \end{bf}
\vspace{1cm}
Yoav Peleg\footnote{This work is supported by the NSF grant
PHY 88-04561 and by a Fishbach Fellowship } \\
\vspace{0.4cm}
Physics Department, Brandeis University, Waltham, MA 02254-9110\\
\vspace{0.2cm}
{\em Peleg@Brandeis}\\
\vspace{0.8cm}
           \begin{large}
ABSTRACT\\
            \end{large}
            \end{center}

By analysing the infinite-dimensional midisuperspace of
spherically symmetric dust matter universes, and applying it
to collapsing dust stars, one finds that the general quantum
state is a bound state. This leads to a discrete spectrum.
To an outside observer, the geometry is static
if the initial radius of the collapsing star is smaller then
the Schwarzschild radius. In that case the discrete spectrum
implies Bekenstein area quantization: the area of the black hole
is an integer multiple of the Planck area. Knowing the
microscopic (quantum) states, we suggest a microscopic
interpretation of the thermodynamics of black holes: by calculating the
degeneracy of the quantum states forming a black hole, one
gets the Bekenstein-Hawking entropy (the entropy is proportional
to the surface area of the black hole). All other
thermodynamical quantities can be derived by using the
standard definitions.

\newpage

\section{Introduction}
Recently a simple minisuperspace describing the Oppenheimer -
Snyder (OS) collapsing star was found [1]. The semiclassical wave
function of that model (e.g. the semiclassical
solution of the Wheeler-DeWitt equation)
is a bound state. This leads to quantization conditions. The
corresponding (Bohr-Sommerfeld) quantization condition can be
written in the form
   \begin{equation}
F(M,R_{0}) = \hbar \left( n + \frac{1}{2} \right) ~~,~~
n = 0,1,2,...
   \end{equation}
where $M$ is the mass of the collapsing star, $R_{0}$ is its
initial radius, and $F(M,R_{0})$ is a function of $M$ and
$R_{0}$ to be given later [1]. For fixed $R_{0}$, (1.1) implies
mass quantization.

The idea of mass quantization
is an old one. Using general arguments
from quantum mechanics (of adiabatic variables) and general
relativity, Bekenstein [2] got the black hole
area quantization condition
   \begin{equation}
M_{ir}^{2} = \frac{1}{2} \hbar n ~~,~~ n = 1,2,3,...
   \end{equation}
where $M_{ir}$ is the irreducible mass of the black hole, which
is related to its surface area , $ {\cal A}
= 16 \pi M_{ir}^{2} $ . This discrete spectrum can be related
to the thermodynamic properties of black holes [3].

We will show that in the case of black holes, one can get (1.2)
from (1.1), which is correct for any OS star.
Beside the quantization condition (1.2), one can find in this
explicit model also the general
quantum states (the solutions of the Wheeler-DeWitt equation)
forming this star. By that one can hope to understand
better not just the quantization conditions,
but also the thermodynamical properties of black holes.
This is the purpose of this paper.

While it has been known for two decades, there are still some open
questions concerning the entropy of black holes.
The classical considerations [4], which use the
analogy between some geometrical properties of black holes
and thermodynamics, gives the ``laws of black hole
thermodynamics", from which one can get the entropy.
Semiclassical considerations [5], on the other hand,
use the (formal) path integrals (approach)
to find the partition function, and then the entropy. But both
approaches do not use basic statistical mechanics reasoning,
namely, finding the entropy by calculating the number of different
``microscopic states" that correspond to the same ``macroscopic state"
that we call a black hole. This ``missing link" is very important,
because it requires the
understanding of {\em microscopic} states forming
a (macroscopic) black hole. Those
microscopic states are the quantum states, so their importance
to the understanding of black holes thermodynamics is obvious.
The OS model gives a one dimensional minisuperspace,
in which of course one cannot hope to get a degeneracy that
will give a nonzero entropy. So one must extend the model.
One such an extension is the inclusion of inhomogeneous (spherically
symmetric dust) distributions. This was done a long time ago by
Lund [6]. Lund used the dust matter as a ``clock" and then fixed
the gauge completely, reducing the constrained matter-gravity
theory to an unconstrained one.
We will use Lund's infinite dimensional ``midisuperspace", and
apply it to the collapsing star case.
Though infinite dimensional, Lund's midisuperspace shares some
resemblance with the OS model, so one can analyse it
in the same manner and find the quantum
states that correspond to a classical black hole.
Knowing the microscopic (quantum) states, one can calculate
their degeneracy, and find the entropy of the black hole. Then
by using the standard thermodynamical definitions one can
get the other thermodynamical quantities (e.g. the temperature).

We will consider both the static ``eternal black hole picture" [7],
and the dynamic Hawking evaporation one [8]. They both can be
studied in our framework, and the results that we get are
in agreement with the known ones.

In this work we use the {\em semiclassical} approximation
only. This is for two reasons: first, the OS model as well as
Lund's one, are correct only semiclassicaly. And second, we
use Einstein gravity (coupled to matter) which should be
(at least) a good approximation semiclassicaly.

We use geometrical units $G = c = 1$.

The paper is organized as follows: in chapter 2 we describe
the OS model, solve (semiclassicaly) the corresponding
Wheller-DeWitt equation, and get the mass and area quantization
conditions. In chapter 3 we describe Lund's midisuperspace and
find the (semiclassical) solutions to the Wheeler-DeWitt equation.
In chapter 4 we describe the midisuperspace of a collapsing star,
and find the quantum states forming this star (black hole).
In chapter 5 we study the thermodynamical properties of black
holes in this framework, and chapter 6 presents some concluding
remarks.

\vspace{1cm}
\section{The OS Model}
\subsection{The OS Minisuperspace}
\setcounter{equation}{0}
In 1939 Oppenheimer and Snyder [9] found a very simple solution
(of Einstein gravity couple to dust matter)
describing a collapse of a spherically symmetric homogeneous dust star.
In their solution the Schwarzschild exterior is smoothly connected
to the interior region, which is a slice of a Friedmann universe.

The interior region is described by the Friedmann line element
   \begin{equation}
d s^{2} = -N^{2}(t) dt^{2} + a^{2}(t) [ d{\chi}^{2} +
sin^{2}{\chi} d {\Omega^{2}_{2}} ]
   \end{equation}
The range of $\chi$ is $0 \leq \chi \leq \chi_{0}$ , where
$\chi_{0} \leq \pi/2$ . At $\chi = \chi_{0}$ the interior is
matched to the exterior Schwarzschild solution. If $M$ and $R_{0}$
are the mass and initial radius of the star, then the matching
conditions are
   \begin{eqnarray}
M &=& \frac{1}{2} a_{0} sin^{3}\chi_{0} \nonumber \\
R_{0} &=& a_{0} sin \chi_{0}
   \end{eqnarray}
where $a_{0}$ is the initial Friedmann radius.

The gravitational Lagrangian may be split into its interior and
exterior parts,
   \begin{equation}
L_{G} = 4\pi \int_{0}^{\chi_{0}} sin^{2}\chi d \chi
\left[ \frac{3a}{N}{\dot{a}}^{2} - 3Na \right] +
\int_{r \geq r_{s}} \sqrt{-g}R d^{3}x
   \end{equation}
where $r_{s}$ is the surface radius of the collapsing star.

The matter Lagrangian is
   \begin{equation}
L_{M} = - 8\pi \int \sqrt{-g} \rho U^{\mu}U_{\mu} d^{3}x
   \end{equation}
where $\rho$ is the density of the star and $U_{\mu}$ is the
four-velocity of the matter particles. Energy momentum
conservation, $\nabla_{\mu} T^{\mu \nu} = 0$, implies
$\rho = \rho_{0}/a^{3}$, where $\rho_{0}$ is a constant to
be determined by the initial conditions. The OS model
requires $\rho_{0} = 3a_{0}/8\pi$. So using (2.3),(2.4) and
$ U^{\mu}U_{\mu} = -1 $ we get
the total Lagrangian
   \begin{equation}
L = L_{G} + L_{M} = 12 \pi \int_{0}^{\chi_{0}} sin^{2}\chi \left[
\frac{a \dot{a}^{2}}{N} - N (a - a_{0}) \right] + \int_{r>r_{s}}
\sqrt{- g} R d^{3}x
   \end{equation}

The Hamiltonian corresponding to (2.5) is
   \begin{equation}
H = N \left[ \frac{1}{4\alpha_{0} a} P_{a}^{2}
+ \alpha_{0} (a - a_{0}) \right]
+ \int_{r>r_{s}} {\cal H} d^{3} x
   \end{equation}
where $ P_{a} = \partial L / \partial \dot{a} $, and $
\alpha_{0} = 12 \pi\int_{0}^{\chi_{0}} sin^{2}\chi d \chi $ .
Because the classical solution for $ r > r_{s} $ is the Schwarzschild
space-time, for which $ R_{Sch.}=0 $, only the first term in (2.5)
(or (2.6)) will contribute to the semiclassical dynamics\footnote{Using
the path integral approach, the semiclassical wave function is
   \begin{eqnarray*}
\psi_{_{WKB}} = A exp[i S_{Class.}/\hbar ]
   \end{eqnarray*}
and we see that only the first term in (2.5) will contribute.}.

The Wheeler-DeWitt equation is the quantum version of the
classical Hamiltonian constraint, $\partial H / \partial N
= 0 $ , in the coordinate representation: $ |\Psi> = \psi(a)
{}~,~ P_{a} = -i\hbar \partial / \partial a $ . Using (2.6) we
get the ``Schr\"{o}dinger equation"
   \begin{equation}
\left( - \hbar^{2} \frac{d^{2}}{d a^{2}} + V(a) \right) \psi(a) = 0
   \end{equation}
where $ V(a) = 4\alpha_{0}^{2} a(a - a_{0}) $ . If we define
$ x \equiv a - a_{0}/2 $, we get
   \begin{equation}
\left( - \hbar^{2} \frac{d^{2}}{d x^{2}} + \frac{1}{4}
\omega^{2} x^{2} \right) \psi(x) = E \psi(x)
   \end{equation}
where $ \omega = 4\alpha_{0} $ and $ E = \alpha_{0}^{2} a_{0}^{2} $.
As we can see from (2.8), $\psi $ describes an harmonic oscillator
(with mass $m = 1/2$). So the semiclassical wave function
describes a one-dimensional harmonic oscillator, which is of
course a bound state.

As in the Hartle-Hawking case [10], the solution of (2.8)
describes a superposition of two ``universes"\footnote{
The solution can be written as
   \begin{eqnarray*}
\psi(x) = A (e^{ip(x)/\hbar} + e^{-ip(x)/\hbar})
   \end{eqnarray*}
where $ p(x) = \int_{x_{0}}^{x} \sqrt{|V(x')|}dx' $.}, one that
collapsses to form a black hole, and one that expands, a white hole.

\vspace{0.5cm}
\subsection{Mass and Area Quantization}
Because the wave function of the OS model describes
a bound state, the spectrum is quantized.
Semiclassically we should use the
Bohr-Sommerfeld quantization condition, but in the case of an
harmonic oscillator it is exact,
   \begin{equation}
E(n) = \hbar \omega \left( n + \frac{1}{2} \right)
{}~~,~~ n = 0,1,2,...
   \end{equation}
Using the definition of $E$ and $\omega$ (see below (2.8)) we have
   \begin{equation}
\frac{1}{4} \alpha_{0} a_{0}^{2} = \hbar \left( n +
\frac{1}{2} \right)
   \end{equation}
In [1] we consider only the case $ R_{0} >> 2M $ which corresponds
to the usual cosmological situation. In that case we have
$ \alpha_{0} \simeq 4 \pi \chi_{0}^{3} $, and using (2.2) and (2.10)
we get [1]
   \begin{equation}
M R_{0}^{3}(n) = \frac{1}{2\pi^{2}} \hbar^{2}
{\left( n + \frac{1}{2} \right)}^{2}
   \end{equation}
For fixed initial radius, (2.11) gives mass quantization.

Of course all the above describes a dynamical process: the collapse
of the star. In this work we try to understand the quantum properties
of a ``static black hole," as seen by an outside observer, and
especially to find its entropy and
temperature (as measured by that observer).
For an outside observer, the above picture can be static only if
$ R_{0} \leq 2M $ . In that case the geometry outside the horizon
is always Schwarzschild, which is of course static. The
OS model requires $ R_{0} \geq 2M $ [9], so
only if $ R_{0} = 2M $ this model can describe a static geometry
everywhere outside the horizon.

In the case $ R_{0} = 2M $ we have $ \chi_{0} =
\pi / 2 $, and we get from (2.2) and (2.10)
   \begin{equation}
M^{2}(n) = \frac{1}{3 \pi^{2}} \hbar \left( n +
\frac{1}{2} \right) ~,
   \end{equation}
a surface area quantization:
the surface area of the black hole\footnote{ For a
Schwarzschild black hole $M_{ir} = M$, so ${\cal A} = 16 \pi
M^{2}$.} goes {\em linearly} with the quantum number $n$.

This is Bekenstein's result [2], but we got it by using an
explicit model, and by solving the corresponding Wheeler-DeWitt
equation. The fact that
we use a very simple (and even a non-realistic) model (the
OS star), and still get his results,
suggests that these are quite general.

In Bekenstein's original paper [2] the prefactor for $M^{2}$
was $ \hbar / 2 $ , but further considerations by Mukhanov [3]
suggest that the prefactor should
be $ \hbar ln 2 / 4 \pi $. In our case the prefactor is $ \hbar
/ 3 \pi^{2} $. This prefactor is a model dependent,
and the best that we
can hope (using our simple model\footnote{ For example,
we take only $R_{0} = 2M$. A more reasonable model should
take some average between $R_{0} = 0$ and $R_{0} = 2M$.}) is to
get the same order of magnitude. This is in fact what we got.

In the OS model, the (microscopic) state of a black hole with
a (macroscopic) mass $M$, is $|\Psi_{n}>$, where $n$ satisfy (2.12).
So in this model, for each macroscopic state (labeled by the mass $M$,
or the area ${\cal A}$) there is only one microscopic state\footnote{
We have a one dimensional harmonic oscillator, so there is no
degeneracy.} (labeled by the quantum number $n$). So the entropy of
the black hole in the OS model is zero\footnote{The entropy
goes like
   \begin{eqnarray*}
S \sim ln(\hbox{number of microscopic states}) = ln(1) = 0
   \end{eqnarray*}}. This is because the OS minisuperspace
is ``too small" (one dimensional). If we want to understand
the thermodynamics of black holes, we must extend the model.
In the next chapter we describe a
much bigger midisuperspace (an infinite dimensional one), which
will turn out to be
much more appropriate for studying black holes thermodynamics.

\vspace{1cm}
\section{Lund's Midisuperspace}
\subsection{The ADM Reduction Process}
\setcounter{equation}{0}
For a global hyperbolic space-time, $M = R \times \Sigma^{(3)}$,
one can use the ADM splitting [11], and write the Hamiltonian of
a dust matter coupled to Einstein gravity in the form [6]
   \begin{equation}
H = \int d^{3}x dt \left[ N ( {\cal H}^{0} + {\cal E} )
+ N_{i} ( {\cal H}^{i} + {\cal P}^{i} ) \right]
   \end{equation}
where $N$ and $N_{i}$ are the lapse function and shift vector
respectively. ${\cal H}^{0}$ and ${\cal H}^{i}$ are the
gravitational super-hamiltonian and super-momentum
   \begin{eqnarray}
{\cal H}^{0} &=& h^{1/2} \left[ \pi^{ij}\pi_{ij} - \frac{1}{2}
\pi^{2} - h R^{(\Sigma)} \right] \\
{\cal H}^{i} &=& -2 h^{1/2} D_{j} \pi^{ij}
   \end{eqnarray}
where $h_{ij}$ is the induced 3-metric on $\Sigma^{(3)}$
($ h = det(h_{ij})$ ), and $\pi^{ij}$ its conjugate momenta
($ \pi = {\pi^{i}}_{i}$ ). ${\cal E}$ and ${\cal P}^{i}$ are
the dust matter Hamiltonian and momentum respectively
   \begin{eqnarray}
{\cal E} &=& h^{1/2} n^{\mu} n^{\nu} T_{\mu \nu} \\
{\cal P}^{i} &=& h^{1/2} n^{\mu} h^{ij} T_{\mu j}
   \end{eqnarray}
where $n^{\mu}$ is a unit normal vector to the hypersurface
$\Sigma^{(3)}$, and $T_{\mu \nu}$ is (using (2.4))
   \begin{equation}
T_{\mu \nu} = 16 \pi \rho U_{\mu} U_{\nu}
   \end{equation}
If we define the scalar field $\phi$ ,
   \begin{equation}
U_{\mu} \equiv \nabla_{\mu} \phi
   \end{equation}
we can treat it as a dynamical variable (and $\rho$, which is
not a dynamical variable, will be a function of $h_{ij},~\pi^{ij},
{}~\phi,~\nabla_{\mu} \phi $, to be determined later.).

In the spherically symmetric case one can use the
$(R,\theta,\phi)$-coordinates  on $\Sigma^{(3)}$ in which
   \begin{equation}
ds_{(3)}^{2} = e^{2\mu} dR^{2} + e^{2\lambda} d{\Omega}^{2}_{2}
   \end{equation}
where $\mu$ and $\lambda$ are functions of $t$ and $R$, and
$d\Omega^{2}_{2}$ is the volume element in $S^{2}$.
If we define $\pi_{\mu}$ and $\pi_{\lambda}$ such that
   \begin{equation}
\pi^{ij} = \mbox{diag}\left( \frac{1}{2} e^{-2\mu}\pi_{\mu},~
\frac{1}{4} e^{-2\lambda}\pi_{\lambda},~\frac{1}{4} e^{-2\lambda}
sin^{-2}\theta \pi_{\lambda} \right)
   \end{equation}
we get
   \begin{eqnarray}
{\cal H}^{0} &=& e^{-(\mu + 2\lambda)} \left( \pi_{\mu}^{2} / 8
- \pi_{\mu} \pi_{\lambda} / 4 - 2 e^{2(\mu + 2\lambda)}
\left[ e^{-2\lambda} - \right. \right. \nonumber \\
& & \left. \left. e^{-2\mu} \left( 2 \lambda" -2\lambda' \mu'
+ 3 (\lambda')^{2} \right) \right] \right) \\
{\cal H}^{R} &=& -e^{-2\mu} ( \pi_{\mu}' - \mu' \pi_{\mu} -
\lambda' \pi_{\lambda} ) \\
{\cal E} &=& {\left( 16 \pi \rho h^{1/2} \right)}^{-1} p_{\phi}^{2}\\
{\cal P}^{R} &=& p_{\phi} h^{11} \phi'
   \end{eqnarray}
where prime denote differentiation with respect to $R$, and
$p_{\phi} = \partial L / \partial \dot{\phi} = -16 \pi N h^{1/2}
U^{0} \rho $. We see that if we choose the coordinates for which
$ N^{R} = 0$, and using $U^{\mu} U_{\mu} = -1$, we get $ \rho
= {(16 \pi h^{1/2})}^{-1} (1 + h^{11} (\phi')^{2})^{-1/2} p_{\phi}$ ,
so from (3.12) we get
   \begin{equation}
{\cal E} = {\left( 1 + h^{11} (\phi')^{2} \right)}^{\frac{1}{2}}
p_{\phi}
   \end{equation}
and we see that ${\cal E}$ goes {\em linearly} with $p_{\phi}$.
This suggests that we can use $\phi$ as a time variable. And
indeed taking $\phi = -t ~,~ N^{R}=0 $ gives (using (3.10)-(3.13))
the known general solutions [12].

To complete the gauge fixing (the reduction process)
one must choose also the $R$-coordinate.
{}From the equations of motion (derive from (3.10)-(3.13))
one can get that $\lambda' e^{\lambda -\mu}$ is
a function of $R$ only, so one can choose
   \begin{equation}
R = \lambda' e^{\lambda - \mu}
   \end{equation}
Now the conjugate momenta are
   \begin{eqnarray}
\pi_{R} &=& -{\left( \lambda' e^{\lambda-\mu} \right)}^{-1}
\pi_{\mu} \\
\bar{\pi}_{\lambda} &=& \pi_{\lambda} - e^{\lambda}
{\left( (\lambda')^{-1} e^{-\lambda} \pi_{\mu} \right)}'
   \end{eqnarray}
Using (3.11) and (3.13), the supermomentum constraint,
$ {\cal H}^{R} + {\cal P}^{R} = 0 $, is now
   \begin{equation}
\pi_{R} + \bar{\pi}_{\lambda} \lambda' + p_{\phi} \phi' = 0
   \end{equation}
After solving the constraints, one ends up with the reduced
Lagrangian (or Hamiltonian)
   \begin{equation}
S_{red} = 4\pi \int dt dR (\pi_{y}\dot{y} - {\cal H}_{ADM} )
   \end{equation}
where
   \begin{eqnarray}
y &=& 8 e^{\lambda} \\
\pi_{y} &=& \frac{1}{8} e^{-\lambda} \bar{\pi}_{\lambda}
   \end{eqnarray}
and
   \begin{equation}
{\cal H}_{ADM} = R^{2} \left( \frac{1}{y} \pi_{y}^{2} +
{(2R)}^{-2} ( R^{-2} -1 ) y \right)
   \end{equation}
So we end up with ${\infty}^{1}$ unconstraint degrees of freedom,
the $y(r)$ field, with the Hamiltonian (3.22). The space of all
$y(r)$-field solutions is what we call ``Lund's midisuperspace".

\vspace{0.5cm}
\subsection{Quantum States}
The reduced Hamiltonian is
   \begin{equation}
H = 4\pi \int {\cal H}_{ADM} dR = 4\pi \int R^{2} \left(
\frac{1}{y} \pi_{y}^{2} + {(2R)}^{-2} ( R^{-2} - 1 ) y \right) dR
   \end{equation}
We use the coordinate representation (in Lund's midisuperspace)
   \begin{eqnarray}
\hat{y} &=& y \\
\hat{\pi}_{y} &=& \frac{\hbar}{i} \frac{\delta}{\delta y}
   \end{eqnarray}
So the corresponding Schr\"{o}dinger equation (in a different context
we could call it the Wheeler-DeWitt equation, see sec. 2) is
   \begin{equation}
i \hbar \frac{\partial \Psi[y;t]}{\partial t} = \int  \left(
- \frac{\hbar^{2}}{y} \frac{\delta^{2}}{\delta y^{2}} +
\frac{1}{4} (R^{-4} - R^{-2}) y \right) \Psi[y;t] R^{2}dR
   \end{equation}
where $|\Psi>$ is the quantum state, and $\Psi[y;t] = <\Psi|
y(R,t)>$ is the wave functional. In this representation it is
a functional of the field $y(r)$ and a function of time $t$.
As one can see, we use the ``$y {\pi}_{y}$-ordering"\footnote{The field
$y$ is always to the left of its conjugate momenta.} in (3.26),
but different ordering will not change our results, which anyway
are correct only semiclassicaly.

A very important feature of (3.23) is that there are no $R$-derivatives
in $H$. This meens that the infinite degrees of freedom (d.o.f.) are
{\em decoupled}. Let $R_{s}$ be the surface of the dust ball,
then we can divide $R_{s}$ to $N$ equal parts, $R_{k} = \frac{R_{s}}{
N} k ~(k=1,2,...,N)$.
The ``continuum limit" is $N \rightarrow \infty$.
The vector space $\{y(r)\}$ is now
$\{ \vec{y} = (y_{1},y_{2},...,y_{N}) \}$ where $y_{k} = y(R_{k})$,
and the Schr\"{o}dinger
equation (3.26) becomes
   \begin{equation}
i \hbar \frac{\partial \Psi(\vec{y},t)}{\partial t} = \sum_{k=1}^{N}
\left[ - a_{k} \frac{\hbar^{2}}{y_{k}} \frac{\partial^{2}}{\partial
y_{k}^{2}} + b_{k} y_{k} \right] \Psi(\vec{y},t)
   \end{equation}
where $ a_{k} = R_{k}^{2} $ and $ b_{k} = (R_{k}^{-2} - 1)/4 $ are
positive constants ($b_{k}$ is positive because $0 \leq R \leq 1$ [6]).
Because of the decoupling we can write $|\Psi>$ as a direct
product
   \begin{equation}
|\Psi> = |\Psi_{1}> |\Psi_{2}> \cdot \cdot \cdot |\Psi_{N}>
   \end{equation}
and from (3.27) we have now
   \begin{equation}
i \hbar \frac{\partial \Psi_{k}(y_{k},t)}{\partial t} =
\left[ - a_{k} \frac{\hbar^{2}}{y_{k}} \frac{\partial^{2}}{\partial
y_{k}^{2}} + b_{k} y_{k} \right] \Psi_{k}(y_{k},t)
   \end{equation}
where $\Psi_{k}(y_{k},t) = < \Psi_{k}|y_{k}(t) >$.
The ADM Hamiltonian is time-independent so we have ${\cal H}_{ADM}
(y_{k}) = E_{k} = const.$ and
   \begin{equation}
\left[ - a_{k} \frac{\hbar^{2}}{y_{k}} \frac{\partial^{2}}{\partial
y_{k}^{2}} + b_{k} y_{k} \right] \Psi_{k}(y_{k},t) = E_{k}
\Psi_{k}(y_{k})
   \end{equation}
If we define
   \begin{equation}
x_{k} \equiv y_{k} - \frac{1}{2} E_{k}
   \end{equation}
we get the following harmonic oscillator Schrodinger equation
   \begin{equation}
\left( - \frac{\hbar^{2}}{2m_{k}} \frac{\partial^{2}}{\partial
x_{k}^{2}} + \frac{1}{2} m_{k}
\omega^{2}_{k} x_{k}^{2} \right) \Psi_{k}(x_{k}) = \epsilon_{k}
\Psi_{k}(x_{k})
   \end{equation}
where $ m_{k}=1/2a_{k} ~,~ \omega_{k} = \sqrt{8a_{k}b_{k}} $ and
$\epsilon_{k}=m_{k}{(\omega_{k}E_{k})}^{2}/8$. The solutions
of (3.32) are $ |\Psi_{k}> = | n_{k} >$ , where $|n_{k}>$ is a
one-dimensional harmonic oscillator with energy
$ \epsilon_{k} = \hbar \omega_{k} ( n_{k} + 1/2 ) $ , or
   \begin{equation}
\frac{1}{8} a_{k}^{1/2} E_{k}^{2} = \hbar \left( n_{k} + \frac{1}{2}
\right)
   \end{equation}
So the space of quantum states describing this spherically symmetric
dust ``universe" is spanned by
   \begin{equation}
\{ |\Psi_{n_{1},n_{2},...,n_{N}} > = |n_{1}>|n_{2}> \cdot \cdot
\cdot |n_{N}> \}
   \end{equation}
The total energy is
   \begin{equation}
E = \sum_{k=1}^{N} E_{k} = \sum_{k=1}^{N} {\left( 8 \hbar
[2(R_{k}^{-4} - R_{k}^{-2})]^{-1/2} \right)}^{1/2} {(n_{k} + 1/2)}^{1/2}
   \end{equation}
In the $N \rightarrow \infty$ limit, one can have a finite
energy only if one uses the Wick order, so $n_{k} + 1/2$ must be
replaced with $n_{k}$, but we will come to that later.

\vspace{1cm}
\section{Midisuperspace for a Collapsing Star}
\subsection{The Homogeneous Case}
\setcounter{equation}{0}
In the homogeneous case (the OS case), it is convenient to
use (2.1). The 3-metric in the $(R,\theta,\phi)$-coordinates, is
   \begin{equation}
h_{ij} = \hbox{diag}\left( a^{2} {\left( \frac{d \chi}{d R}
\right)}^{2} ,~ a^{2}sin^{2} \chi,~ a^{2}sin^{2}\chi sin^{2} \phi
\right)
   \end{equation}
and from (3.8),(3.15) and (4.1) we get
   \begin{equation}
R = \lambda' e^{\lambda - \rho} = cos \chi
   \end{equation}
We see that $R$ is not a ``usual radial coordinate".
For example the origin ($r=0$) is $R=1$, and $R_{min} =
cos \chi_{0} \geq 0$. From (3.15) one can see that $ R = r_{c}/
r_{e}$ [6], where $r_{c}$ is the circumference radius, and $r_{e}$
is the extrinsic radius of curvature.
Because the horizon is a minimal area, $R(horizon)=0$.
One can get this explicitly in the OS model, because
a static black hole corresponds to
$\chi_{0}=\pi/2$, so $ R(horizon) = cos \chi_{0} = 0 $.
For asymptotically flat space $R$ goes to unity at spatial
infinity. It is convenient to use a different radial
coordinate\footnote{The coordinate $r$ grows with the usual
radial coordinate while $R$ decreases.}
   \begin{equation}
r \equiv sin \chi = {(1 - R^{2})}^{1/2}
   \end{equation}
Now one should replace (3.23) with
   \begin{equation}
H = \int_{0}^{r_{s}} {\cal H}(r) dr =
\int_{0}^{r_{s}} \left( \frac{32 r \sqrt{1-r^{2}}}{3 \pi}
\frac{P_{y}^{2}}{y} + \frac{3 \pi r}{2\sqrt{1-r^{2}}} y \right) dr
   \end{equation}
where $ y = 8e^{\lambda} = 8ar $, and $ P_{y}= 3\pi y \dot{y}/64r\sqrt{
1- r^{2}}$. It is very easy to see that (4.4) is the correct Hamiltonian,
because from (2.6) and (4.4) we have $ P_{y} = 3\pi r P_{a} / 2 \alpha_{0}
\sqrt{1-r^{2}}$, so
   \begin{eqnarray}
H &=& 12 \pi \int_{0}^{sin\chi_{0}} dr \frac{r^{2}}{\sqrt{1-r^{2}}}
\left( \frac{P^{2}_{a}}{4 \alpha_{0}^{2} a} + a \right) \nonumber \\
&=& \frac{P^{2}_{a}}{4 \alpha_{0} a} + \alpha_{0} a
   \end{eqnarray}
which is exactly the gravitational part of (2.6).

The Schr\"{o}dinger equation is
  \begin{equation}
\hat{H} |\Psi> = E |\Psi>
  \end{equation}
In Lund's formalism $E$ is an undefined constant, but in the
collapsing star case, we must impose the matching conditions:
the solution must be smoothly matched to the outside Schwarzschild
space-time. This will constrain $E$. From (2.1) and (2.4) we have
   \begin{equation}
E = 32\pi^{2} \int_{0}^{\chi_{0}} sin^{2}\chi d \chi a^{3} \rho
   \end{equation}
Because $E$ is time independent, we can easily calculate it at
the beginning of the collapse, when the star is at rest. Using
$ a(t=0) = a_{0} $ and $ \rho(t=0)= 3 M / 4\pi R_{0}^{3} $, we get
   \begin{equation}
E = \frac{ 2 \alpha_{0} a_{0}^{3} M}{R_{0}^{3}}
   \end{equation}
Now we can use the matching conditions (2.2) to get
   \begin{equation}
E = \alpha_{0} a_{0}
   \end{equation}
and the Schr\"{o}dinger equation (4.6) is exactly (2.7),and one
can get the results of chapter 2.

Notice that we could get (4.9) from the requirement that
the collapse start ($t=0$) at rest. We will use that in the
inhomogeneous case.

In the homogeneous case the difference between Lund's
midisuperspace and a collapsing star midisuperspace is that
in the latter the energy $E$ is constraint. In the inhomogeneous
case the constraints are more complicated, but one can
deal with them in a similar way.

\vspace{0.5cm}

\subsection{The Inhomogeneous Case}
We saw that (in the homogeneous case) one can use Lund's
formalism, impose the energy condition (4.9), and  get the OS
results. This can be generalized to
the inhomogeneous case.

Let us first go beck to the homogeneous case. Using (4.4)
and (4.9), the Schrodinger equation (4.6) can be written as
   \begin{equation}
\int_{0}^{r_{s}} \left[ \frac{32 r\sqrt{1-r^{2}}}{3\pi}
\frac{P_{y}^{2}}{y}+ \frac{3\pi r}{2\sqrt{1-r^{2}}} (y - y_{0})
\right] \Psi[y] dr = 0
   \end{equation}
where $y_{0}=y(r,t=0)=8a_{0}r$.

Now we can generalized to the inhomogeneous case. The result
(3.23) (or (4.4)) are correct for {\em any} spherically symmetric
dust ball (not just for the homogeneous one). In particular, they
are correct in the case of a collapsing (spherically symmetric
dust) star. If we want the collapse to start ($t=0$) at
rest, which is the generalization of (4.9),
then we must have\footnote{$E=\int_{0}^{r_{s}} {\cal E}
(r) dr$, so ${\cal H}(r) = {\cal E}(r)$, and from (4.4) and (4.10)
we get (4.11).}
   \begin{equation}
{\cal E}(r) = \frac{3 \pi r}{2\sqrt{1-r^{2}}}y_{0}
   \end{equation}
because only in that case $\dot{y}(r,t=0) = 0$. So the form
of (4.10) is quite general: it is correct also in the inhomogeneous
case. The only
difference is that in the inhomogeneous case, the
Friedmann radius, $a$, can be a
function of $r$ too. So as a function of $(r,t)$ the field
solution is different, but it has the same form $y(r,t) =
8a(r,t)r$. The space of all field solutions that describe
a collapsing star is of course a subspace of all the field
solutions (Lund's midisuperspace), but at this stage we do
not need to know the specific restrictions; what is
important is that we can use (4.10).

After making the discretization $ r_{k}=r_{s} k / N
{}~(k=1,2,...,N) $ , we get from (4.10)
   \begin{equation}
\sum_{k=1}^{N}\left( \alpha_{k}\frac{P_{y_{k}}^{2}}{y_{k}}
+ \beta_{k} (y_{k} - y_{k}^{(0)} ) \right) \Psi(\vec{y}) = 0
   \end{equation}
where $ \alpha_{k} = (32 r_{k} \sqrt{1-r_{k}^{2}})/3\pi $ , $ \beta_{k} =
3 \pi r_{k} / 2 \sqrt{1-r_{k}^{2}} $ and $ y_{k}^{(0)} = y_{0}(r_{k}) $.
Using $ \Psi(\vec{y}) = \prod_{k} \Psi_{k}(y_{k}) $ we have
a set of N independent equations
   \begin{equation}
\left( \alpha_{k}\frac{P_{y_{k}}^{2}}{y_{k}}
+ \beta_{k} (y_{k} - y_{k}^{(0)} ) \right) \Psi_{k}(y_{k}) = 0
   \end{equation}
Defining $ x_{k} \equiv y_{k} - y_{k}^{(0)} / 2 $ , we get the
(harmonic oscillator) Schr\"{o}dinger equation
   \begin{equation}
\left( -\frac{\hbar^{2}}{2m_{k}} \frac{\partial^{2}}{
\partial k_{k}^{2}} + \frac{1}{2} m_{k}
\omega^{2}_{k} x^{2}_{k} \right) \Psi_{k}(x_{k}) =
\epsilon_{k} \Psi_{k}(x_{k})
   \end{equation}
where $ m_{k} = 1 / 2 \alpha_{k} $ , $ \omega_{k} = 8 r_{k} $
and $ \epsilon_{k} =  m_{k} \omega_{k}^{2} {(y_{k}^{
(0)})}^{2} / 8 $. The
quantization conditions are
   \begin{equation}
r_{k} m_{k} {(y_{k}^{(0)})}^{2} = \hbar \left( n_{k} + \frac{1}{2}
\right)
   \end{equation}
The total energy is
   \begin{equation}
E = 3 \pi \sum_{k=1}^{N} \frac{r_{k} y_{k}^{(0)}}{2
\sqrt{1-r_{k}^{2}} }
   \end{equation}
and from (4.15) we get
   \begin{equation}
E = \hbar \sum_{k=1}^{N} \Omega_{k} (n_{k} + 1/2)
   \end{equation}
where $\Omega_{k} = 4 / a_{k}^{(0)} $, $(a_{k}^{(0)} = a(r_{k},
t=0))$.

So a quantum state describing a collapsing dust star, starting
at rest, can be written as
   \begin{equation}
|\Psi(\mbox{star})> = |n_{1}>|n_{2}> \cdot \cdot \cdot |n_{N}>
   \end{equation}
where $|n_{k}>$ is a one dimensional harmonic oscillator
(exited to the level $n_{k}$) with frequency $\omega_{k}=8r_{k}$.
Each $|n_{k}>$ and so $|\Psi>$ are bound states,
and we end up with quantization conditions.

In the homogeneous case $ y_{k}^{(0)} = 8 a_{0} r_{k} $ ,
($ a_{k}^{(0)} = a_{0} $) and from (4.17) we get
   \begin{equation}
E_{hom} = \frac{4 \hbar}{ a_{0} } \sum_{k} (n_{k}
+ 1/2 )
   \end{equation}
and using (4.9) we have
   \begin{equation}
\frac{1}{4} \alpha_{0} a_{0}^{2} = \hbar \sum_{k} (n_{k} + 1/2)
   \end{equation}
which is (2.10) (remember that in the homogeneous case there is
only one d.o.f. so $N=k=1$) .

In the general inhomogeneous case, $\Omega_{k}$ is not
$k$-independent, but we can use ``mean field" reasoning
and write (4.17) as
   \begin{equation}
E = \hbar \sum_{k=1}^{N} \Omega_{k} (n_{k} + 1/2) =
\hbar <\Omega> \sum_{k=1}^{N} (n_{k} + 1/2)
   \end{equation}
One can use (4.21) as a definition of $<\Omega>$, which is
the ``average" of $\Omega_{k}$.

The results (4.18) and (4.21) are correct for any collapsing
spheriacly symmetric dust star, which start at rest. But in
the case of a static black hole, one can relate $E$ and
$<\Omega>$ to the mass of the black hole. In the homogeneous
case we have $ a_{0} = 2M_{bh} $ so from (4.9) we have $E = 6
\pi^{2} M_{bh}$, and from (4.17) $<\Omega> = \Omega_{k} =
{(2M_{bh})}^{-1}$. In the inhomogeneous case this can be generalized
by dimensional arguments to
   \begin{equation}
E \sim \frac{1}{<\Omega >} \sim M_{bh}
   \end{equation}
wher $\sim$ denote equality up to a constant of order unity.
Now using (4.21) and (4.22) we get
   \begin{equation}
M^{2}_{bh} \sim \hbar \sum_{k=1}^{N} (n_{k} + 1/2)
\equiv \hbar (n_{tot} + 1/2)
   \end{equation}
Which is the generalization of (2.12) to the general
inhomogeneous case, in agreement with Bekenstein's result.

\vspace{1cm}
\section{Black Hole Thermodynamics}
\subsection{Entropy}
\setcounter{equation}{0}
Using the standard statistical mechanics definition,
the entropy of a (macroscopic) black hole is
   \begin{equation}
S_{bh} =  \mbox{ln} [{\cal N}_{bh}(M)]
   \end{equation}
where we choose $k_{B}$ to be one, and ${\cal N}_{
bh}(M)$ is the number of microscopic states that correspond
to the same macroscopic state (a black hole)
with mass $M$. The microscopic states are (4.18), and for a
given $M$ (or a given $E$), ${\cal N}_{bh}(M)$ is the number
of different $|\Psi(\mbox{star})>$-states, that satisfy (4.21)
(or (4.23)).

Consider first the limit $N \rightarrow \infty$:
according to (4.21) the energy, $E$, will
be finite only if
we use the Wick order. Then
if only a finite number of d.o.f. are excited, $E$ will be finite.
In that case there are infinitely many other d.o.f. that are in their
ground states, and we face two (probably related) problems:
first, our semiclassical approximation is very bad when most of
the d.o.f. are in their ground state. Second, ${\cal N}_{bh}$
is infinite (there are infinitely many ways to choose a
finite number of exited d.o.f. from an infinite number of total
d.o.f.) so the entropy will diverge.
As a matter of fact, the limit $N \rightarrow \infty$ is
questionable. For example (2.10) is as much a radius
quantization condition as it is a mass quantization condition.
This means that one cannot simply divide $a_{0}$ infinitely many
times. This is clearly a quantum gravitational issue.
But there is a way to avoid it:
If we choose {\em not} to use the Wick order,
then if $E$ is finite, $N$ must be finite too.
We see that (2.12) or its generalization (4.23) provide us
with a natural cutoff\footnote{In our geometrical units $M_{P} =
l_{P} = \hbar^{1/2}$.}
$N_{max} \sim (M/M_{P})^{2}$, which for a classical black
hole is a big, but still finite.

Of course $ 1 \leq N \leq N_{max} $. In a sense, $N$ describe
the amount of inhomogenity. For $N=1$ we have the homogeneous
case, for $N=2$ the ``almost" homogeneous one, and so on until
$N=N_{max}$ which describe the general inhomogeneous star.
We have then
   \begin{equation}
{\cal N}_{bh}(M) = \sum_{N=1}^{N_{max}} {\cal N}_{bh}
(N,M)
   \end{equation}
where ${\cal N}_{bh}(N,M)$ is the corresponding number
for a specific $N$. Our semiclassical approximation is
good as long as $N$ is much smaller then $N_{max}$, but we
will see that the contribution to (5.2) from $ N > N_{max}/2 $
is the same as from $ N < N_{max}/2 $. So at least we have
a good estimate to (5.2).

It is easy to see that the $N$'s that
will contribute to (5.2) must satisfy
$N = N_{max} - 2j ~,~ j = 0,1,...,(N_{max}-1)/2 $.
Let us start from $N=N_{max}$. In that case
we have only one state (4.18), $|\Psi > = |0>_{1} |0>_{2} ...
|0>_{N_{max}}$ . Next we consider $N=N_{max}-2$, it is
easy to see that they are $N_{max}-2$ states\footnote{The
oscilators are distinguishable because they have different
frequencies, $\omega_{k} = 8 r_{k}$. Or in other words:
different $k$'s correspond to different shells
which are distinguishable because they have different radii.},
$ |\Psi> = |0>_{1}..|1>_{k}..|0>_{N_{max}-2} $. In a similar
way, we have for any $N=N_{max}-2j$
   \begin{equation}
{\cal N}_{bh}(N,M) = C^{j}_{N_{max}-1-j}
   \end{equation}
where $C^{k}_{m}$ are the binomial coefficients. So
   \begin{equation}
{\cal N}_{bh}(M) = \sum_{j=0}^{(N_{max}-1)/2}
C^{j}_{N_{max}-1-j}
   \end{equation}
Though there is no known analytic expresion for (5.4) [13], it is
elementary to show numericaly that
   \begin{equation}
\sum_{j=0}^{n} C^{j}_{2n-j} \simeq \mbox{exp}(
0.962 n - 0.320 )
   \end{equation}
In our case we have $n=(N_{max}-1)/2 \sim (M/M_{P})^{2}$, so
   \begin{equation}
{\cal N}_{bh}(M) \sim \mbox{exp}\left( C \frac{M^{2}}{M_{P}^{2}}
\right)
   \end{equation}
where $C$ is a constant of order unity.
And we find (using (5.1)) that the entropy of the black hole is
   \begin{equation}
S_{bh} =  C \frac{M^{2}}{M_{P}^{2}} + S_{0}
   \end{equation}
The entropy is proportional to the surface area of the
black hole, or equivalently, it goes linearly with the quantum
number $n_{tot}$, see (4.23), in agreement with the
Bekenstein-Hawking entropy.

We could use a different approach to calculate the entropy:
One can use the Wick order, so (4.22) is replaced with
   \begin{equation}
E = \hbar <\Omega> \sum_{k=1}^{N} n_{k}
   \end{equation}
but still take a finite $N_{max}$:
{}From (2.10) and $R_{0} \sim a_{0}$ we have that
$ \Delta R_{0} \geq \hbar / R_{0} $, and from $ R_{0} = a_{0} r_{s} $ we
get $ (\Delta R_{0})_{min} = a_{0} r_{s} / N_{max} $. So in the
case of black holes, $r_{s}=1$ and $ R_{0} \sim M $ , we
have $N_{max} \sim R^{2}_{0} / \hbar \sim M^{2} / M^{2}_{P}$.
In this case  $N$ can take all the integer values between $N_{max}$ and
unity; the degeneracy is
   \begin{equation}
{\cal N}_{bh}(M) = \sum_{N=1}^{N_{max}-1} C^{N}_{N_{max}-1}
= 2^{N_{max} - 1}
   \end{equation}
Now the entropy is
   \begin{equation}
S_{bh} = ln2 \frac{M^{2}}{M^{2}_{P}} + \tilde{S}_{0}
   \end{equation}
We see that (5.7) and (5.10) have the same form, the entropy is
proportional to the surface area of the black hole, but the
prefactors are different. One should get the correct prefactor
in the full exact model.

Another thing to notice is that the degeneracy (and so the
entropy) of the gravitational d.o.f. is very similar to the
degeneracy of other field d.o.f. [14], so it is tempting to
think that the gravitational d.o.f. that we use are more
appropriate for a unified scheme.

\vspace{0.5cm}
\subsection{Temperature}
Using the standard thermodynamical definition of the
temperature we have
  \begin{equation}
T_{bh}^{-1} = \frac{ \partial S_{bh} }{ \partial E_{bh}}
  \end{equation}
We have $E_{bh} \sim M$, and using (5.6) we get
   \begin{equation}
T_{bh} \sim \frac{M_{P}^{2}}{M}
   \end{equation}
in agreement with the Hawking temperature [8].

One may argue that the microscopic states (4.18)
are microscopic both to a freely falling observer
(``Kruskal observer") and to an outside observer (``Schwarzschild
observer"). This means that (5.1) should be the same for Kruskal
observers as well. Then the entropy (5.7), and temperature (5.12),
are the same for both the
Kruskal and Schwarzschild observers. This contradicts
the known results that a thermal Schwarzschild state corresponds
to a zero Kruskal temperature [7,14]. But remember
that though the microscopic states (4.18) are the same for
both observers, the {\em macroscopic states} are quite different.
For a Schwarzschild observer there is an horizon, and from the
no-hair theorems, there is only one macroscopic quantity (in
the Schwarzschild case) by which one can determine the state. This
is of course the mass $M$ of the star. In that case the degeneracy
is exactly what we get from (4.21), and indeed we have (5.7) and (5.12).
On the other hand, for a Kruskal observer, there is no horizon,
and the macroscopic state is determine by an {\em infinite} number
of macroscopic quantities. For example in our model, we have a
global hyperbolic space-time, so a classical solution is determined
by the initial data\footnote{The question of classical (and of
course quantum) observables in gravity
is an open one, but {\em in principle}
one should be able to determine those quantities.}
$(y(r,t=0),\dot{y}(r,t=0))$. In our case we
have $\dot{y}(r,t=0) = 0$, so a classical solution is determine
by $y(r,t=0)$, or equivalently by all the moments
$ P_{n} = \int_{0}^{r_{s}} r^{n} y(r,t=0) dr $, which are
macroscopic quantities. This means that all the $\Omega_{k}$'s
in (4.23) are determined, and (at least semiclassically) the state (4.18)
is determined {\em completely} by the macroscopic state. This means
that for a Kruskal observer there is no degeneracy, and the
entropy and temperature vanish.

The entropy (5.1) is sometimes called ``entangeled entropy",
but we think that (at least in the case of black holes) it
should be consistent with the thermal entropy. The way to
check this is to couple the system to other fields.
When we couple the gravitational d.o.f. to other fields, we have
the following picture: The fields are in ``equilibrium" with
the gravitational d.o.f. (the black hole). According to a
Schwarzschild observer, it is a thermal equilibrium with
the temperature (5.12), and according to a Kruskal observer it is
a zero temperature situation. This is a static scenario,
in agreement with the ``eternal black hole" picture [7],
and with the Thermo-field approach [15].

One can consider also a dynamical situation: a black hole
creation and evaporation. This will be done in the next subsection.

\vspace{0.5cm}
\subsection{Hawking Evaporation}
So far we have studied only a static picture as seen by a Schwarzschild
observer, in which a black hole is in a thermal state in equilibrium
with the outside region. But one can use our formalism to study
also the dynamical process of Hawking evaporation. Using the
semiclassical adiabatic arguments, one assumes that at each time the
star is in a state (4.18), and the Hawking radiation is the result
of a transition between a level $|\Psi(n_{tot})>$ (see (4.23)) to one of the
closest levels, $|\Psi(n_{tot}-1)>$ [3]. Using energy conservation
and (4.23), the radiation frequency satisfies
   \begin{equation}
\hbar \omega_{rad} = \Delta M(n_{tot},n_{tot}-1) \sim
\frac{M_{P}^{2}}{M} ~.
   \end{equation}
On the other hand the temperature is proportional to the
radiating energy (frequency), so we have
   \begin{equation}
T_{H} \sim \hbar \omega_{rad} = \Delta M \sim
\frac{M^{2}_{P}}{M}
   \end{equation}
in agreement with (5.12), and with Hawking results.

In this dynamical situation, one can calculate the lifetime
of the level $|\Psi(n_{tot})>$ [3]. This should be finite, because
there is an interaction with the vacuum state of the radiation fields.
Now this is not the Kruskal vacuum\footnote{Known as the
Hartle-Hawking [16], or Israel [15] vacuum.}
(like in the eternal black
hole case). The vacuum state is now the Unruh vacuum [17].
This lifetime can be estimated to be proportional to the
inverse of the imaginary part of the effective action (in the
Unruh vacuum), and one get the mass rate [3]
   \begin{equation}
\frac{d M}{d t} \sim \frac{T}{\Delta M} \sim \frac{
M^{2}_{P}}{M^{2}}
   \end{equation}
in agreement with Hawking results, which assume a black body
radiation rate.

If we ``extrapolate" our results to the quantum region
($n \sim 1$), we can say that there should be a ``quantum
remnant" of mass $M_{rem} \sim M_{P}$ at the end of the
Hawking evaporation [18]. But this is pure speculation
because we ignore back-reaction as well as strong
quantum effects in our model.

\vspace{1cm}
\section{Concluding Remarks}
\setcounter{equation}{0}
In this work we used the canonical quantization approach
of spherically symmetric dust matter universes, first given
by Lund, and apply it to the case of collapsing stars and
black holes. The quantum states describing those universes
are bound states and one gets a discrete spectrum.

First let us consider some of the physical consequences of
the quantized spectrum.
One may question the collapse process itself, because if the
collapsing star must satisfy (2.11) for example, then in the
space of all masses and initial radii, only a set of measure
zero satisfy it, so maybe most of the stars will not
collapse at all? This is not the case because though the
mass is a constant (by energy conservation), $R_{0}$ (or in
the general case, all the other geometrical quantities) can
fluctuate, and one must calculate $ \Delta R_{0} / R_{0} $.
If this is a very small number, then the collapse is possible
in a general situation. Using (2.11) we have
   \begin{equation}
\frac{\Delta R_{0}}{R_{0}} \sim \frac{\hbar^{2} n}{
M R_{0}^{3} } \sim {\left( \frac{M_{P}}{M} \right)}^{1/2}
{\left( \frac{l_{P}}{R_{0}} \right)}^{3/2} \sim \frac{1}{n}
   \end{equation}
which for astronomical objects is a very small number. For
example in the case of our sun, we have $ \Delta R_{\odot}
/ R_{\odot} \sim 10^{-100} $.
This means that for astronomical objects, the mass quantization
cannot affect the classical collapse process. On the other hand,
if we ``extrapolate" our results to Plankc size objects, then
the collapse itself may be affected by the quantization
conditions. This may be another reason to consider stable
Planck size objects (black holes)?

The effect of the mass quantization on the Hawking radiation
spectrum, will be mainly on very large wavelengths, $ \lambda
\geq M_{bh}$. The black hole cannot radiate or absorb radiation
with $\lambda > M_{bh}$ because it correspond to $\Delta M$
smaller than the distance between nearest levels. For astronomical
objects this effect will be hard to detect.

So we see that for ``classical objects" (astronomical stars),
for which $n >> 1$, the correspondence principle works,
and the quantum effects are negligible.

It is also quite easy to recover the classical laws of
black hole thermodynamics: using (4.23) and ${\cal A}
= 16 \pi M^{2}$  we get the first law of
(Schwarzschild) black hole thermodynamics
   \begin{equation}
\delta M = \frac{\partial M}{\partial n} \delta n \sim
\frac{M^{2}_{P}}{M} \delta {\cal A}
   \end{equation}
And because $M_{bh} \sim E_{bh}$ we have $S_{bh} \sim {\cal A}$ and
$T_{bh} \sim M^{2}_{P} / M_{bh}$.

The second law is just (5.7) with the fact that $\Delta S_{bh}
\geq 0$ for an isolated system, while the generalized second
law is $\Delta S_{tot} \geq 0$, where $ S_{tot} = S_{bh} + S $.

In the case of spherically symmetric dust matter, the infinite
number of gravitational degrees of freedom decoupled, and
each shell of dust moves independently. It is possible to
choose the coordinates and the field variables such that each
shell is an harmonic oscillator. This
is a simple generalization of the homogeneous Oppenheimer-
Snyder model.

In the case of black holes, the discrete spectrum gives
the Bekenstein area quantization: the area of the black hole
is an integer number times the Planck area.

It is very easy to calculate the degeneracy of this system
(of independent oscillators), and from it to get the entropy
of the black hole. The results agree with the known
Bekenstein-Hawking entropy: the entropy is proportional to
the surface area of the black hole. Then one can use the
standard thermodynamic definitions to get all the other
thermodynamic quantities (e.g. Hawking temperature).

It seems surprising that our simple model (of spherically
symmetric dust matter) gives ``enough" degeneracy, and the
correct Bekenstein-Hawking entropy. One might think that in
the general case (no symmetry, and a general
matter field) the degeneracy will be much bigger, and so also
the entropy, which would contradict known results.
But this is not necessarily the case. One should remember
that we have been able to quantize the dust system, because
we could fix the gauge completely, which means that we correctly
choose the coordinate system and solved the
constraints. A consistent quantization of the general
case (if it exists) may be achieved by a ``free field
representation", which will be a generalization of our
independent harmonic oscillators. If this is the case,
then the general degeneracy will be quite similar to what
we have in the dust case, as will the black hole thermodynamics.
Maybe this is one thing that we can learn from our simple
model\footnote{Though infinite-dimensional.}.
It should be interesting to study some extensions of
our model, and to see if our results will survive.
Another thing that may be interesting to study in more detail,
is the interaction between the geometrical
degrees of freedom that we use and other fields.

\vspace{1cm}
{\bf Aknowledgment}\\
I would like to thank S. Deser and G. `t Hooft for very
helpful discassions.

\newpage

{\bf REFERENCES}\\
\begin{enumerate}
\item Y.Peleg, ``The Wave Function of a Collapsing Star and
Quantization Conditions", Brandeis University Report No. BRX-TH-
342, (1993) hepth@xxx/9303169
\item J.D.Bekenstein, {\em Lett.Nuov.Cimen.}{\bf 11}, 476 (1974)
\item V.F.Mukhanov, {\em JEPT Lett.}{\bf 44},63 (1986) \\
J.Garcia-Bellido, ``Quantum Black Holes", Stanford
University Report No. SU-ITP-93-4 (1993)
\item J.W.Bardeen, B.Carter and S.W.Hawking, {\em Comm.Math.Phys.}
{\bf 31}, 161 (1973)
\item G.W.Gibbons and S.W.Hawking, {\em Phys.Rev.}{\bf D15}, 2738
(1977)
\item F.Lund, {\em Phys.Rev.}{\bf D8}, 3253; 4229 (1973)
\item G.W.Gibbons and M.J.Perry, {\em Proc.R.Soc.London},{\bf A358},
467 (1978)
\item S.W.Hawking, {\em Comm.Math.Phys.}{\bf 43},199 (1975)
\item J.R.Oppenheimer and H.Snyder, {\em Phys.Rev.}{\bf 56},455
(1939)\\
C.W.Misner, K.S.Thorne and J.A.Wheeler, {\em Graviataion} (Freeman,
San Fransisco, 1973)
\item J.B.Hartle and S.W.Hawking, {\em Phys.Rev.}{\bf D28},2960
(1983)
\item R.Arnowitt, S.Deser and C.W.Misner, {\em Ann.Phys.(N.Y)}
{\bf 33}, 88 (1965); ``The Dynamics of General Relativity", in
{\em Gravitation: An Introduction to Current Research}, ed. L.
Witten (Wiley, New-York, 1962)
\item R.C.Tolman, {\em Proc.Nat.Acad.Sci.}{\bf 20}, 169 (1934)\\
L.D.Landau and L.M.Lifshitz, ``The Classical Theory of Fields",
(Pergamon, Oxford, 1962)
\item G.P.Egorychev, ``Integral Representation and the
Computation of Combinatorial Sums", (American Mathematical
Society, Providence, 1984)
\item M.Srednicki, ``Entropy and Area", Berkeley Report No.
LBL-33754, hepth@xxx/9303048 (1993)
\item W.Israel, {\em Phys.Lett.}{\bf A57}, 107 (1976) \\
Y.Takahashi and H.Umezawa, {\em Collec.Phen.}{\bf 2},55, (1975)
\item J.B.Hartle and S.W.Hawking, {\em Phys.Rev.}{\bf D13},
2188 (1976)
\item W.G.Unruh, {\em Phys.Rev.}{\bf D14}, 870, (1976)
\item Y.Aharonov, A.Casher and S.Nussinov, {\em Phys.Lett.}{\bf
B191}, 51 (1987)

\end{enumerate}

\end{document}